\def\beqar{\begin{eqnarray}}
\def\eeqar#1{\end{eqnarray}}

\documentstyle[ichep,12pt]{article}

\title{\normalsize MSW IMPLICATIONS OF
SOLAR NEUTRINO EXPERIMENTS\thanks{This work is supported in part by the U. S.
Department of Energy grant DE-FG05-92ER40691.}}

\author{S. P. Rosen\\ College of Science\\ The University of Texas at
        Arlington\\ Arlington, Texas 76019-0047\\ U. S. A.}

\begin{document}
\finalcopy
\maketitle

\vskip -2ex
\abstract{I discuss the implications for future solar neutrino experiments of
the most recent gallium data in the context of the MSW mechanism. At the low
energy end of the solar neutrino spectrum we need to measure the $^7$Be
component directly; and at the high energy end, we need precise measurements of
the shape of the spectrum.  }

\vskip-1pc

\onehead{INTRODUCTION}

John Bahcall concluded his talk by telling us that the Standard Solar Model
(SSM) is in good shape and that the observed deficit of solar neutrinos must be
due either to serious flaws in the existing experiments or to new physics
associated with non-zero neutrino mass. I am going to adopt the second
alternative, namely that the experiments are broadly correct and that the
deficit is caused by the elegant Mikheyev-Smirnov-Wolfenstein (MSW) matter
enhancement of neutrino oscillations. The new SAGE and GALLEX results that we
heard here today from Drs.\,Vignaud and Gavrin have a clear interpretation in
terms of the MSW effect and point to specific expectations for the next
generation of experiments. It is these expectations that I shall discuss in
this talk.

I shall begin with a brief review of the MSW solution to the solar neutrino
problem, then discuss the interpretation of the latest gallium data and develop
the implications for \break

\newpage\noindent
new experiments now under construction. I shall
distinguish between the `low energy' end of the solar neutrino spectrum and
the `high energy' end: the low energy end consists of $pp$ neutrinos with
energies up to 430 keV and $^7$Be neutrinos with an energy of 860 keV. The high
energy part of the spectrum consists of $^8$B neutrinos with energies all the
way up to 14 MeV. Future experiments include BOREXINO, which is designed to
look at the $^7$Be neutrinos and SNO, Super Kamiokande, and ICARUS which will
detect the high energy neutrinos.

\onehead{BRIEF REVIEW OF MSW}

In the standard electroweak model, neutrinos of all flavors can scatter from
electrons via the exchange of a neutral $Z^0$ boson; the electron-neutrino can,
in addition, scatter through the exchange of a charged $W$ boson. This unique
diagram gives the electron-neutrino a different refractive index, or effective
mass, in matter and it also yields a much larger cross-section for the process
of neutrino--electron scattering than does the neutral current diagram.

The time development equation for MSW matter oscillations between the electron
flavor and another active flavor $x$ can be written as:
\beqar
i\frac{d}{dt} \pmatrix{a_e\cr a_x\cr}  = \pmatrix{X&Y\cr Y&Z\cr}
                                              \pmatrix{a_e\cr a_x\cr}
\eeqar{1}
where $a_e$, $a_x$ are the probability amplitudes for $\nu_e$ to survive, and
to oscillate into flavor $x$ respectively. The matter Hamiltonian is given by:
\begin{eqnarray*}
\noalign{\vskip -2.5ex}
       X &=& \frac{m^2_1 c^2 + m^2_2 s^2}{2E} + \sqrt{2}\, G_F N_e \\
       Z &=& \frac{m^2_1 s^2 + m^2_2 c^2}{2E} \\
       Y &=& \frac{m^2_2 - m^2_1}{2E} cs  = \frac{\Delta m^2}{2E} cs
\end{eqnarray*}
where $c = \cos \theta$, and $s = \sin \theta$.
With the right electron density $N_e$ in the matrix element $X$ we
can tune the Hamiltonian to maximal mixing, {\it i.e.} $X = Z$, or
\beqar
   \sqrt{2}\, G_F N_e =  \frac{\Delta m^2}{2E} \cos 2\theta
\eeqar{5}
and thereby obtain enhanced conversion of the electron-neutrino to
flavor $x$.

There are two types of solution to eq.~(1): the adiabatic solution in which the
instantaneous eigenvectors change slowly as the neutrino travels through the
sun, and the nonadiabatic solution, which involves a transition between
eigenvectors. The criterion for validity of the adiabatic solution is:
\beqar
  \frac{\sin^2 2\theta}{\cos 2\theta}\frac{\Delta m^2}{E} \gg
  \frac{1}{N}\frac{dN}{dR}
\eeqar{6}
Taking the scale height of the solar density on the right-hand side
from the standard solar model, we can write this as
\beqar
E \ll 3.5\times 10^8 \frac{\sin^2 2\theta}{\cos 2\theta} \Delta m^2\ \rm MeV,
\eeqar{7}
where $\Delta m^2$ is measured in eV$^2$. The electron-neutrino survival
probability at Earth is
\beqar
  P_{ad}(\nu_e \rightarrow \nu_e) &= & \frac{1}{2}
  \left(1 - \frac{T \cos 2\theta}{\sqrt{T^2 + \sin^2 2\theta}}\right) \nonumber
  \eeqar{}
  \beqar
  T &= & 1.53\times 10^{-5}\frac{E}{\Delta m^2}\frac{N}{N_c} - \cos 2\theta.
\eeqar{8}
where $N_c$ is the density at the core of the sun. Key points on the curve of
$P_{ad}$ as function of energy are shown in Table 1.

\begin{table}
\begin{center}
\renewcommand{\arraystretch}{1.2}
\begin{tabular}{ccc}
\multicolumn{3}{l}{Table 1. $P_{ad}(\nu_e \to \nu_e)$ as a function}\\
\multicolumn{3}{l}{\hbox to 48pt{\hfil}of neutrino energy}   \\ \hline\hline
$E_{\nu}$ & $T$             &  $P(\nu_e \to \nu_e)$          \\ \hline
0         & $-\cos 2\theta$ & $(1 - \frac12 \sin^2 2\theta)$ \\
$E_0$     & 0               & $\frac12$                      \\
$2E_0$    & $+\cos 2\theta$ &$\frac12 \sin^2 2\theta$        \\
$\infty$  &$\infty$         &$\sin^2 \theta$                 \\ \hline\hline
\end{tabular}
\end{center}
\end{table}

For small mixing angles, the location of the enhancement energy $E_0$ and the
rapid fall in survival probability between the origin and $2E_0$ are
insensitive to the to the precise value of the mixing angle; the enhancement
energy is, however, directly proportional to $\Delta m^2$.

Equation (5?) shows that the adiabatic survival probability depends upon the
birthplace of the neutrino in the sun. Since the $pp$ and $^7$Be neutrinos are
produced over a broader region of the solar interior than $^8$B neutrinos, this
will contribute to differences in behavior of low and high energy neutrinos. To
take this effect into account, we integrate $P_{ad}$ over the production
region:
\beqar
  \langle P_{ad} \rangle &= &\int \phi (r) P_{ad}(\nu_e \to \nu_e)dr
\eeqar{9}
where $\phi(r)$ is the fraction of neutrinos produced at a radius $r$ inside
the  sun, with $\int\phi(r)dr = 1$.  Its values for every branch of the solar
neutrino spectrum are tabulated by Bahcall and Ulrich.

The nonadiabatic solution is valid when the inequality of eq.~(3) is reversed
and it yields a very simple form for the electron-neutrino survival
probability at Earth:
\beqar
   P_{nonad}(\nu_e \rightarrow \nu_e) &= & \exp(-C/E),\nonumber
   \eeqar{}
   \beqar
   C &= & \frac{\pi \Delta m^2 \sin^2 2\theta}{4\cos 2\theta}
          \left( \frac{1}{N}\frac{dN}{dR} \right)^{-1} \nonumber \\
   &=& 2.75 \times 10^8 \Delta m^2 \frac{sin^2 2\theta}{\cos 2\theta}~\rm MeV.
\eeqar{10}
This expression is independent of the point of birth of the neutrino and needs
no integration over the production zone. From the fit to the $^{37}$Cl data, we
find that
\beqar
  \Delta m^2 \frac{sin^2 2\theta}{\cos 2\theta}\approx 3\times10^{-8}~\rm eV^2,
\eeqar{11}
and hence $C$ is approximately 9 MeV.

For small mixing angles, these solutions meet at a point approximately[RefPet]
where the nonadiabatic probability of eq.~(10) equals the asymptotic value
$\sin^2 2\theta$ of the adiabatic survival probability (see Fig.1). The
neutrino energy versus mass$^2$ difference at this point is
\beqar
   \left( \frac{E_j}{\Delta m^2} \right) =  \left( \frac{- 2.75 \times 10^8
       sin^2 2\theta}{\cos 2\theta \ln(\sin^2 \theta)} \right).
\eeqar{12}
Imposing the $^{37}$Cl nonadiabatic constraint of eq.~(11), we obtain
the energy at the joining point
\beqar
\noalign{\vskip -2.5ex}
   E_j = \frac{- 8.25}{\ln (\sin^2 \theta)} =  1.3~\rm MeV
\eeqar{13}
for $\sin^2 2\theta = 0.007$. This formula means that not all solar $\nu_e$
will have a nonadiabatic survival probability: those with
energies greater than some value in the neighborhood of $E_j$
will obey $P_{nonad}$, but those with energies {\em less than} $E_j$ will
survive according to the {\em adiabatic} probability in eq.~(8). The behavior
of the electron-neutrino survival probability in this case is shown in
Figs.~1 and 2.

\onehead{THE GALLIUM RESULTS}

We heard from Vignaud that the GALLEX signal is approximately $80 \pm 20$ SNU,
and from Gavrin that the latest SAGE result is about $60 \pm 20$ SNU. Both
measurements are significantly larger than the original SAGE result which
clearly favored the nonadiabatic MSW solution over the adiabatic and the large
angle ones. As pointed out by the GALLEX collaboration, the present results are
still consistent with a nonadiabatic solution, but now admit a large angle
solution as an alternative possibility. Thus we need to find means by which to
distinguish between them.

Let us consider how the nonadiabatic solution for the
$^{37}$Cl experiment can accommodate relatively large signals in the $^{71}$Ga
experiment. As originally pointed out by Rosen and Gelb, the gallium
signal varies along the nonadiabatic line described by eq.~(11) above from very
small values near the center to much larger ones at either end. At the upper
end in particular, where $\sin^2 2\theta$ is relatively small and $\Delta m^2$
large, the expected signal comes close to the SSM prediction. This happens
because the electron-neutrino survival probability decreases with decreasing
energy from 14 MeV until about 1 MeV and then begins to increase up the
adiabatic curve as the energy decreases
further. The slope of this curve (see Figs. 1,2), causes the lower
energy $pp$ neutrinos to have a significantly greater survival probability than
the $^7$Be neutrinos.

The relationship between these two low energy components of the spectrum is
different in the large angle solution for $^{37}$Cl, which is  described by the
adiabatic formula of eq.~(8). High energy $^8$B neutrinos fall in the
asymptotic
region with  survival probability $\sin^2 \theta$. Low energy neutrinos are in
the region  where, as energy decreases, the survival probability increases to
$(1 - \frac{1}{2}\sin^2 2\theta)$; the $pp$ neutrinos still have a greater
survival probability than the $^7$Be ones, but the difference is not as large
as in the small angle case. Thus, one way of making the choice between the two
solutions is to study the relative survival probabilities of $pp$ and $^7$Be
neutrinos. We now examine the situation in more detail.

\onehead{NONADIABATIC, SMALL ANGLE SOLUTION}

The nonadiabatic, small angle solution given by GALLEX consists of a small
region around the point
\beqar
\sin^2 2\theta = 0.007,~~\Delta m^2 = 6\times 10^{-6}.
\eeqar{14}
$^8$B neutrinos survive nonadiabatically  and low energy ones adiabatically.
For a density of 70 gm/cc, which is typical of the solar production zones for
$pp$ and $^7$Be neutrinos, the key points of the adiabatic probability in Table
1 take on the numerical values in the second and third columns of Table~2.

\begin{table}
\begin{center}
\renewcommand{\arraystretch}{1.2}
\begin{tabular}{c@{\hspace{2em}}cc@{\hspace{2em}}cc}
\multicolumn{5}{l}{Table 2. $P(\nu_e \to \nu_e)$ as a function of}\\
\noalign{\vskip-0.5ex}
\multicolumn{5}{l}{\hbox to 48pt{\hfil}neutrino energy (in keV).}\\
\noalign{\vskip1ex}
\hline\hline
&\multicolumn{2}{l}{small angle}&\multicolumn{2}{l}{large angle}\\
Point    & $E_{\nu}$ & $P(\nu_e)$ & $E_{\nu}$ & $P(\nu_e)$\\ \hline
origin   & 0        & 0.997 & 0        & 0.7 \\
$E_0$    & 564      & 1/2   & 475      & 1/2 \\
$2E_0$   & 1,128    & 0.003 & 950      & 0.3 \\
$\infty$ & $\infty$ & 0.002 & $\infty$ & 0.2 \\ \hline\hline
\end{tabular}
\end{center}
\end{table}

Since $pp$ neutrinos are well below 564 keV in energy and $^7$Be ones well
above, the former have a survival probability much greater than 1/2 and the
latter much less. The same holds true for most densities encountered in the
production zone for low energy neutrinos and so we expect the gallium signal to
be predominantly composed of $pp$  neutrinos. Calculations of $\langle P_{ad}
\rangle$ (see eq.~(9)) for the 860 keV $^7$Be neutrino and for a `typical' $pp$
of 300 keV are plotted as functions  of $\Delta m^2$ in Fig.~3; for $\Delta m^2
= 6\times 10^{-6}$, $^7$Be  neutrinos have about a 5\% survival probability and
$pp$ neutrinos have close  to $100\%$. Therefore the $pp$ neutrinos contribute
about 90\% of the observed  gallium signal in the small angle solution.

\onehead{LARGE ANGLE SOLUTION}

The large angle GALLEX solution is centered upon the point
\beqar
   \sin^2 2\theta = 0.6,~~\Delta m^2 = 8\times 10^{-6}.
\eeqar{15}
Again the adiabatic formula is relevant for the low energy neutrinos and, with
with a density of 70 gm/cc, the key points are shown in the fourth and fifth
columns of Table 2.

The $pp$ and $^7$Be neutrinos are still on opposite sides of the enhancement
energy, but the difference in their survival probabilities is not as marked as
in the small angle case. The integrated probability $\langle P_{ad} \rangle$
(see eq.~(9)) for the 860 keV $^7$Be neutrino and for a `typical' $pp$ of 300
keV maintain this feature, as can be seen in the plots of Fig.~4. For
$\Delta m^2 = 8\times 10^{-6}$, the $^7$Be neutrinos have a 30\% survival
probability and $pp$ neutrinos have a 60\% one. Thus the $pp$ neutrinos
contribute about half of the observed gallium signal and the other branches of
the spectrum the other half in the large angle solution.

\Onehead{IMPLICATIONS FOR EXPERIMENTS}{AT THE LOW ENERGY END}

It is clear that new experiments at the low energy end of the spectrum must
attempt to measure the $pp$ and $^7$Be neutrinos either by themselves alone, or
in some combination distinct from that of existing experiments. One
proposal, the BOREXINO experiment, plans to measure the $^7$Be neutrinos by the
process of neutrino--electron scattering and in the SSM it expects to see 47
recoil electrons per day in the energy range of 250 to 663 keV. The upper
energy represents the kinematic limit for incident $^7$Be neutrinos and the
lower energy is almost at the kinematic limit for incident $pp$ neutrinos.
Contributions from other branches of the solar spectrum are negligible. Thus
this experiment would give us a clear shot at the $^7$Be neutrinos.

I assume that the electron neutrino oscillates under the MSW effect into
another active flavor. The
electron-neutrino itself will scatter from an electron via a coherent
combination of charged- and neutral-currents and the other flavors via
neutral-currents alone; thus the former will have a cross-section roughly 6
times larger than the latter. If the survival probability for $^7$Be
electron-neutrinos is $P_7$, then the number of events predicted per day will
be
\beqar
R_7 = (0.79 P_7 + 0.21) \times 47.
\eeqar{16}
By the time BOREXINO comes on the air in a few years, the errors on the gallium
experiments should be considerably smaller than they are today; if nothing
else, the statistics should be significantly improved. I therefore reduce the
current errors on $\sin^2 2\theta$ and $\Delta m^2$ by a factor of two and
assume that the central values will not change significantly.

In the small angle solution, survival probabilities are not sensitive to the
mixing angle and our reduced errors indicate that $\Delta m^2$ should lie
between 4 and 6 $\times 10^{-6}$. From Fig.~4, it can be seen that $P_7$ lies
between 0 and $5\%$ and that the expected rate in BOREXINO would be:
\beqar
    10 \leq R_7 ~{\rm (small)} \leq 12 ~\rm events/day.
\eeqar{17}
In the large angle solution, $\Delta m^2$ lies between 4 and 6 $\times 10^{-6}$
and $P_7$ between 0.3 and 0.5 (see Fig.~4). The expected rate in this case
turns out to be
\beqar
    21 \leq R_7 ~{\rm (large)} \leq 28 ~\rm events/day.
\eeqar{18}
Comparing the two expectations for $R_7$, we see that BOREXINO should be able
to make a clear distinction between the two MSW solutions for existing solar
neutrino experiments.

For $pp$ neutrinos, the small angle solution predicts a survival probability of
85\% or more, while the large angle one predicts about 65\%. Thus we will need
reasonable accurate measurements of the $pp$ neutrinos to distinguish the  two
cases directly.

\Onehead{IMPLICATIONS FOR EXPERIMENTS}{AT THE HIGH ENERGY END}

At the high energy end of the spectrum, the small angle solution yields an
exponential survival probability as in eq.~(10) with the parameter $C$ between
9  and 11 MeV. The large angle solution yields a constant probability of
$\sin^2 \theta$ and equal to 0.2 for $\sin^2 2\theta$ = 0.6. Our question is
whether the next generation of experiments will accumulate sufficient
statistics with sufficient accuracy to distinguish between these two survival
probabilities.

In order to make this distinction, it is necessary to measure the spectrum of
electron-neutrinos arriving at Earth. When the survival probability is flat,
the spectral {\em shape} will be exactly as predicted by the SSM, but the
overall normalisation will be reduced; when the probability is a function of
neutrino energy, the spectral shape will be distorted. For the exponential
formula of eq.~(10), the largest distortions come at the lower energies and so
it is important for the next generation of experiments to push thresholds to as
low an energy as possible.

Some idea of the difficulties involved in measuring spectral shapes can be
obtained from early work of Bahcall, Gelb and Rosen on neutrino--electron
scattering and a recent analysis of the Kamiokande II data by Kwong and Rosen.
It is clear that differences in the spectral shapes of the neutrinos themselves
become muted in the convolutions that lead to the observed recoil electron
spectrum, especially when the threshold on electron energy is as high as 7.5
MeV. Were it 5, or better still 3 Mev, then it might be possible to use the
observed electron spectrum directly, but at higher thresholds we need to find
clever tricks.

One such trick, originally used by the Kamiokande II collaboration, is to
compare the observed recoil electron spectrum with that predicted by the SSM on
a point-by-point basis. This serves to emphasise any differences between the
effects of different neutrino survival probabilities. In particular, the ratio
is flat as a function of energy for the large angle MSW solution, and has a
positive slope for the small angle one. Unfortunately, the errors on the
present Kamiokande II data are too large to yield any definite conclusion.
If the larger part of the errors is statistical, then several thousand events
would be needed to reduce them to a useful level for our purposes. Super
Kamiokande, collecting events at the rate of 10 per day, could collect a useful
sample in one year and, depending upon its size and threshold, so could ICARUS.

SNO would take several years for neutrino--electron scattering, but it will
have
a much more powerful tool in neutrino-induced disintegration of the deuteron.
Through the neutral-current process it can establish whether or not
oscillations into active flavors are taking place, and through measurements of
the electron spectrum in charged-current disintegration, it can measure the
spectrum of electron-neutrinos arriving at Earth. Here too, the distinctions
between different cases are muted by convolutions of spectra and reaction
cross-sections, and it may be necessary to resort to the same kind of trick as
in neutrino--electron scattering.

\onehead{CONCLUSION}

Matter oscillations and the MSW Effect still provide a viable and attractive
solution of the solar neutrino problem, but we must wait until the next
generation of experiments before we can know for sure that this is so. As the
gallium experiments acquire more data and reduce errors, our confidence in the
outcome will increase or decrease according as the central value of the signal
decreases or increases. The next five years should be very interesting.

\end{document}